\documentclass[sigconf]{acmart}
\acmConference[MSR 2024]{21st International Conference on Mining Software Repositories}{April 2024}{Lisbon, Portugal}

\usepackage{graphicx} 
\usepackage{xcolor}
\usepackage{xspace}

\title{Predicting the Impact of Crashes Across Release Channels}

\author{Suhaib Mujahid}
\affiliation{%
  \institution{Mozilla Corporation}
  \city{Montreal}
  \country{Canada}}
\email{smujahid@mozilla.com}

\author{Diego Elias Costa}
\affiliation{%
  \institution{Concordia University}
  \city{Montreal}
  \country{Canada}}
\email{diego.costa@concordia.ca}

\author{Marco Castelluccio}
\affiliation{%
  \institution{Mozilla Corporation}
  \city{London}
  \country{UK}}
\email{mcastelluccio@mozilla.com}

\begin{document}

\begin{abstract}
Software maintenance faces a persistent challenge with crash bugs, especially across diverse release channels catering to distinct user bases. Nightly builds, favoured by enthusiasts, often reveal crashes that are cheaper to fix but may differ significantly from those in stable releases. In this paper, we emphasize the need for a data-driven solution to predict the impact of crashes happening on nightly channels once they are released to stable channels. We also list the challenges that need to be considered when approaching this problem.
\end{abstract}
\maketitle



\section{Background and Motivation}


Modern software maintenance is characterized by a fast and continuous cycle of releases. 
Within each release, developers evolve their system by releasing updates to meet new user demands, but need also to work tirelessly in fixing bugs to ensure the reliability and quality of their software. 
Crashing bugs, among the many types of bugs, are one of the most impactful for software reliability~\cite{AnLe:CrashingBugs:15}. 
These bugs often result in abrupt and unintended termination of software applications, causing user frustration and system outages, leading to potentially significant costs.

To cope with the fast release cycles, contemporary development practices use a variety of release channels~\cite{DaCosta:RapicRelease:16}. 
These release channels, such as Nightly builds, Beta versions, and Stable Releases, cater to different segments of the user base~\cite{whattrainisitnow}. 
The Nightly channel is typically used by enthusiasts and early adopters who are interested in the latest features, albeit with a higher tolerance for bugs and instability. 
In contrast, the Stable Release channel is designed for the broader public, prioritizing reliability and consistency.
This tiered approach enables developers to balance innovation and reliability, by receiving early feedback on new features on rapid-release channels (e.g., Nightly), and progressively refining their software for more stable channels~\cite{Castelluccio:2019:RapidRelease,Clark:RapidReleaseSecurity:14}. 

Crashes identified in more frequent release channels (e.g., Nightly) are cheaper to fix and impact fewer (more risk-taking) users. 
However, the user profile of crash bugs that occur in the Nightly channel may differ significantly from the crashes that occur in the more stable channels.
For example, Nightly users, being enthusiasts, usually have more powerful machines than the average Stable Release user.
Thus, developers may decide not to prioritize crashing bugs that rarely occur in the Nightly channel, only to later realize their significant impact on the more stable release channels, affecting the larger user base of the software~\cite{Kim_TSE2011}.
For example, a network bug that caused hangs in the socket thread and made the browser unresponsive was not initially detected in the Beta or Nightly versions. However, the bug became apparent following the move to the Stable Release channel. It was found that the bug was also present in the Nightly and Beta versions, but it was overlooked due to the small volume of affected users using these versions. The bug only became noticeable when it was moved to the Stable Release channel~\cite{bug1749910}.

We shall discuss the complex decision-making aspects that release managers use today, and, more importantly, the opportunities for developing a data-driven solution for \textbf{predicting the impact of crashing bugs in stable releases.}
A potential solution could involve analyzing crash volumes and their characteristics across various release channels.


\section{Challenges}
The task of correlating reported crashes across different release channels presents significant challenges to researchers and practitioners alike. Below, we outline the main challenges.

\textbf{Fixed Bugs:} Developers are continuously working to fix bugs and crashes. As a result, a crash that occurs in a Nightly version may be fixed after moving that version to the Beta channel.
As some users are still using an older version, new crash reports will continue to come in for both Nightly and Beta channels even after the fix has been implemented. However, the volume of these reports on Nightly will be smaller. This makes it somewhat difficult to correlate the volume of reports across channels. One potential solution could be to use the build ID to mitigate this issue, at the cost of limiting the amount of available data in some cases.

\textbf{Feature toggles:} It is common to have experimental features that are enabled in Nightly and limited to a subset of users on Beta or Stable Release channels using feature toggles~\cite{Rahman:FeatureToggles:16}. However, enabling such features only on a subset of users can lead to an inaccurate reflection of the true crash volume on wider channels if the feature is fully enabled.

\textbf{Gradual rollout:} The process of distributing a version update on Beta or Stable Release channels occurs gradually. Initially, only a subset of the users receive the update, and if everything goes well, the new version will eventually be available to all users. This ensures a smoother rollout and a better user experience. However, during this period, the volume of crashes may be underestimated.

\section{Datasets}

The dataset for the crash reports from Firefox users can be publicly assessed through the Mozilla Crash Reports API~\cite{CrashStats}. The API allows querying of crash reports submitted in the last six months. Additionally, information from the bug tracking systems associated with the actionable crashes can be accessed through the Bugzilla WebService API~\cite{BugzillaAPI}.
Alternatively, some datasets can be downloaded through BugBug, Mozilla's platform for machine learning projects on software engineering~\cite{BugBug_docs_data,Castelluccio_bugbug_2021}.

Firefox developers use the crash signature to group related crashes in the same bucket. When investigating a crash to be fixed, a bug report is filed on Bugzilla and linked to the crashes through one or more crash signatures.
We shall discuss how to use the available data to help foment more research on the topic of predicting the impact of crashing bugs across release channels.

\bibliographystyle{ACM-Reference-Format}
\bibliography{bibliography}


\begin{thebibliography}{12}


\ifx \showCODEN    \undefined \def \showCODEN     #1{\unskip}     \fi
\ifx \showDOI      \undefined \def \showDOI       #1{#1}\fi
\ifx \showISBNx    \undefined \def \showISBNx     #1{\unskip}     \fi
\ifx \showISBNxiii \undefined \def \showISBNxiii  #1{\unskip}     \fi
\ifx \showISSN     \undefined \def \showISSN      #1{\unskip}     \fi
\ifx \showLCCN     \undefined \def \showLCCN      #1{\unskip}     \fi
\ifx \shownote     \undefined \def \shownote      #1{#1}          \fi
\ifx \showarticletitle \undefined \def \showarticletitle #1{#1}   \fi
\ifx \showURL      \undefined \def \showURL       {\relax}        \fi
\providecommand\bibfield[2]{#2}
\providecommand\bibinfo[2]{#2}
\providecommand\natexlab[1]{#1}
\providecommand\showeprint[2][]{arXiv:#2}

\bibitem[An and Khomh(2015)]%
        {AnLe:CrashingBugs:15}
\bibfield{author}{\bibinfo{person}{Le An} {and} \bibinfo{person}{Foutse Khomh}.} \bibinfo{year}{2015}\natexlab{}.
\newblock \showarticletitle{An Empirical Study of Highly Impactful Bugs in Mozilla Projects}. In \bibinfo{booktitle}{\emph{2015 IEEE International Conference on Software Quality, Reliability and Security}}. \bibinfo{pages}{262--271}.
\newblock
\urldef\tempurl%
\url{https://doi.org/10.1109/QRS.2015.45}
\showDOI{\tempurl}


\bibitem[Castelluccio(2023)]%
        {Castelluccio_bugbug_2021}
\bibfield{author}{\bibinfo{person}{Marco Castelluccio}.} \bibinfo{year}{2023}\natexlab{}.
\newblock \bibinfo{booktitle}{\emph{{BugBug}}}.
\newblock
\urldef\tempurl%
\url{https://doi.org/10.5281/zenodo.4911345}
\showDOI{\tempurl}


\bibitem[Castelluccio et~al\mbox{.}(2019)]%
        {Castelluccio:2019:RapidRelease}
\bibfield{author}{\bibinfo{person}{Marco Castelluccio}, \bibinfo{person}{Le An}, {and} \bibinfo{person}{Foutse Khomh}.} \bibinfo{year}{2019}\natexlab{}.
\newblock \showarticletitle{An Empirical Study of Patch Uplift in Rapid Release Development Pipelines}.
\newblock \bibinfo{journal}{\emph{Empirical Softw. Engg.}} \bibinfo{volume}{24}, \bibinfo{number}{5} (\bibinfo{date}{oct} \bibinfo{year}{2019}), \bibinfo{pages}{3008–3044}.
\newblock
\showISSN{1382-3256}
\urldef\tempurl%
\url{https://doi.org/10.1007/s10664-018-9665-y}
\showDOI{\tempurl}


\bibitem[Clark et~al\mbox{.}(2014)]%
        {Clark:RapidReleaseSecurity:14}
\bibfield{author}{\bibinfo{person}{Sandy Clark}, \bibinfo{person}{Michael Collis}, \bibinfo{person}{Matt Blaze}, {and} \bibinfo{person}{Jonathan~M. Smith}.} \bibinfo{year}{2014}\natexlab{}.
\newblock \showarticletitle{Moving Targets: Security and Rapid-Release in Firefox}. In \bibinfo{booktitle}{\emph{Proceedings of the 2014 ACM SIGSAC Conference on Computer and Communications Security}} (Scottsdale, Arizona, USA) \emph{(\bibinfo{series}{CCS '14})}. \bibinfo{publisher}{Association for Computing Machinery}, \bibinfo{address}{New York, NY, USA}, \bibinfo{pages}{1256–1266}.
\newblock
\showISBNx{9781450329576}
\urldef\tempurl%
\url{https://doi.org/10.1145/2660267.2660320}
\showDOI{\tempurl}


\bibitem[Coman(2022)]%
        {bug1749910}
\bibfield{author}{\bibinfo{person}{Marius Coman}.} \bibinfo{year}{2022}\natexlab{}.
\newblock \bibinfo{title}{Bug 1749910: Hangs in socket thread}.
\newblock
\newblock
\urldef\tempurl%
\url{https://bugzilla.mozilla.org/show_bug.cgi?id=1749910}
\showURL{%
\tempurl}
\newblock
\shownote{[Online; accessed 29. Nov. 2023]}.


\bibitem[da~Costa et~al\mbox{.}(2016)]%
        {DaCosta:RapicRelease:16}
\bibfield{author}{\bibinfo{person}{Daniel~Alencar da Costa}, \bibinfo{person}{Shane McIntosh}, \bibinfo{person}{Uir\'{a} Kulesza}, {and} \bibinfo{person}{Ahmed~E. Hassan}.} \bibinfo{year}{2016}\natexlab{}.
\newblock \showarticletitle{The Impact of Switching to a Rapid Release Cycle on the Integration Delay of Addressed Issues: An Empirical Study of the Mozilla Firefox Project}. In \bibinfo{booktitle}{\emph{Proceedings of the 13th International Conference on Mining Software Repositories}} (Austin, Texas) \emph{(\bibinfo{series}{MSR '16})}. \bibinfo{publisher}{Association for Computing Machinery}, \bibinfo{address}{New York, NY, USA}, \bibinfo{pages}{374–385}.
\newblock
\showISBNx{9781450341868}
\urldef\tempurl%
\url{https://doi.org/10.1145/2901739.2901764}
\showDOI{\tempurl}


\bibitem[Kim et~al\mbox{.}(2011)]%
        {Kim_TSE2011}
\bibfield{author}{\bibinfo{person}{Dongsun Kim}, \bibinfo{person}{Xinming Wang}, \bibinfo{person}{Sunghun Kim}, \bibinfo{person}{Andreas Zeller}, \bibinfo{person}{S.C. Cheung}, {and} \bibinfo{person}{Sooyong Park}.} \bibinfo{year}{2011}\natexlab{}.
\newblock \showarticletitle{Which Crashes Should I Fix First?: Predicting Top Crashes at an Early Stage to Prioritize Debugging Efforts}.
\newblock \bibinfo{journal}{\emph{IEEE Transactions on Software Engineering}} \bibinfo{volume}{37}, \bibinfo{number}{3} (\bibinfo{year}{2011}), \bibinfo{pages}{430--447}.
\newblock
\urldef\tempurl%
\url{https://doi.org/10.1109/TSE.2011.20}
\showDOI{\tempurl}


\bibitem[{Mozilla Corporation}(2023a)]%
        {BugzillaAPI}
\bibfield{author}{\bibinfo{person}{{Mozilla Corporation}}.} \bibinfo{year}{2023}\natexlab{a}.
\newblock \bibinfo{title}{{Bugzilla WebService API Reference}}.
\newblock
\newblock
\urldef\tempurl%
\url{https://bmo.readthedocs.io/en/latest/api/index.html}
\showURL{%
\tempurl}
\newblock
\shownote{[Online; accessed 28. Nov. 2023]}.


\bibitem[{Mozilla Corporation}(2023b)]%
        {CrashStats}
\bibfield{author}{\bibinfo{person}{{Mozilla Corporation}}.} \bibinfo{year}{2023}\natexlab{b}.
\newblock \bibinfo{title}{{Mozilla Crash Stats}}.
\newblock
\newblock
\urldef\tempurl%
\url{https://crash-stats.mozilla.org/documentation}
\showURL{%
\tempurl}
\newblock
\shownote{[Online; accessed 28. Nov. 2023]}.


\bibitem[{Mozilla Corporation}(2023c)]%
        {whattrainisitnow}
\bibfield{author}{\bibinfo{person}{{Mozilla Corporation}}.} \bibinfo{year}{2023}\natexlab{c}.
\newblock \bibinfo{title}{The Release Channels of {Firefox} Desktop}.
\newblock
\newblock
\urldef\tempurl%
\url{https://whattrainisitnow.com}
\showURL{%
\tempurl}
\newblock
\shownote{[Online; accessed 30. Nov. 2023]}.


\bibitem[Mujahid(2023)]%
        {BugBug_docs_data}
\bibfield{author}{\bibinfo{person}{Suhaib Mujahid}.} \bibinfo{year}{2023}\natexlab{}.
\newblock \bibinfo{title}{Downloading Data Using BugBug}.
\newblock
\newblock
\urldef\tempurl%
\url{https://github.com/mozilla/bugbug/blob/master/docs/data.md}
\showURL{%
\tempurl}
\newblock
\shownote{[Online; accessed 30. Nov. 2023]}.


\bibitem[Rahman et~al\mbox{.}(2016)]%
        {Rahman:FeatureToggles:16}
\bibfield{author}{\bibinfo{person}{Md~Tajmilur Rahman}, \bibinfo{person}{Louis-Philippe Querel}, \bibinfo{person}{Peter~C. Rigby}, {and} \bibinfo{person}{Bram Adams}.} \bibinfo{year}{2016}\natexlab{}.
\newblock \showarticletitle{Feature Toggles: Practitioner Practices and a Case Study}. In \bibinfo{booktitle}{\emph{Proceedings of the 13th International Conference on Mining Software Repositories}} (Austin, Texas) \emph{(\bibinfo{series}{MSR '16})}. \bibinfo{publisher}{Association for Computing Machinery}, \bibinfo{address}{New York, NY, USA}, \bibinfo{pages}{201–211}.
\newblock
\showISBNx{9781450341868}
\urldef\tempurl%
\url{https://doi.org/10.1145/2901739.2901745}
\showDOI{\tempurl}


\end{thebibliography}

\end{document}